\documentstyle[aps,preprint,12pt,subeqn,epsf,rotate]{revtex}
\newcommand{\gl}[1]{(\ref{#1})}

\newcommand{\citefigure}[1]{Fig.~\ref{#1}}
\def\gtrless{\raise2.5pt\hbox{$>$}\llap{\lower2.5pt\hbox{$<$}}}
\begin{document}

\title{
Structural relaxation in orthoterphenyl: 
a schematic mode coupling theory model analysis} 
\author{ A.P.~Singh$^a$, G.~ Li$^b$, 
 W.~G{\"o}tze$^a$, M.~Fuchs$^a$, T.~Franosch$^a$ and H.Z.~Cummins$^b$} 
\address{$^a$Physik-Department, Technische Universit{\"a}t M{\"u}nchen, 85747
Garching, Germany;\\
$^b$Department of Physics, City College -- CUNY, New York, NY 10031, USA}
\date{Received  December  1997}
\maketitle

\begin{abstract}
\noindent
Depolarized light scattering spectra of orthoterphenyl showing the emergence of
the structural relaxation below the oscillatory microscopic excitations are
described by solutions of a  schematic mode--coupling--theory model. 
\end{abstract}
\vspace{2cm}
\centerline{PACS numbers: 64.70.Pf, 61.20.Lc}

\vspace{2cm}
{\tt accecpted for publication in J. Non-Cryst. Solids}

\newpage
\section{Introduction}
\noindent
Structural relaxation in glass forming liquids has been studied in great
detail. Typically spectra are measured in the supercooled state for
temperatures $T$ near the glass transition temperature $T_g$ focusing on
frequencies 
lower than, say, 10 MHz. However, it has been shown in many recent experiments
that structural dynamics can be observed also for higher temperatures, even
 above the
melting temperature $T_m$. The relaxation spectra then show up  
in  a frequency
 window just below the band of microscopic excitations, which is located
near and above 1 THz. For example, the depolarized light scattering spectra of
orthoterphenyl (OTP) exhibit a non--exponential $\alpha$--process within the
GHz band for $T= T_m + $50 K  \cite{eins,zwei}. 

The evolution of the 
structural relaxation upon cooling is the theme of the mode--coupling theory
(MCT) of the glass transition \cite{fuenf}. This theory describes the dynamics
of a  liquid by a set of  non--linear equations of motion, which
use the equilibrium structure functions as input. Structural relaxation is
obtained as anomalous dynamics, caused by a bifurcation singularity at some
critical temperature $T_c$, which is located typically between $T_g$ and
$T_m$. Usually, data interpretation within MCT is based on universal results,
obtained by solving the MCT equations asymptotically for $T$ 
close to $T_c$. It was
shown in ref. \cite{eins} that a major part of the OTP--data could be
described by using the universal MCT results for the $\alpha$-- and
$\beta$--process. In the same frame neutron scattering spectra could be
explained as is shown in ref. \cite{vier} and papers quoted therein.

The shape of the $\alpha$--peak  is not a universal feature of structural
relaxation. It depends on the details of the microscopic structure and it is
different for different probing variables \cite{fuenf}. The universal results
of the MCT 
deal with relaxation only, and thus they cannot be used to discuss the
crossover from relaxation at low frequencies to oscillations at higher
frequencies. These shortcomings in the application of universal
MCT results can be
overcome to some extent by the application of schematic MCT models, as was
shown recently in a discussion of the light scattering spectra of glycerol
\cite{drei}. Such models truncate the microscopic MCT equations 
 so that the universal features remain intact and that
e.g. some of the oscillatory dynamics can still be described. Thereby, one can
get solutions for the analysis of spectra which cover huge windows and which do
not rely on asymptotic formulas.
In this paper the MCT model of ref. \cite{drei} shall be used to interpret the
light scattering spectra of OTP ($T_m=$ 329 K, $T_g=$ 244 K) from
ref. \cite{eins} for frequencies between 0.04 GHz and 400 GHz and temperatures
between 225 K and 395 K.

\section{Schematic mode--coupling model}
\noindent
The model to be used   assumes that the structural
relaxation of density fluctuations
can be modeled by a single oscillator equation with a retarded
friction function $m(t)$:
\begin{equation}
\partial_{t}^{2}{\Phi}(t)+\nu\partial_{t}{\Phi}(t)+\Omega^{2}\Phi(t)+
\Omega^{2}\int_{0}^{t}~ m(t-t')~ \partial_{t'}{\Phi}(t')~ {\mbox{d}}t' = 0.
\label{e1}
\end{equation}
The time--dependent friction $m(t)$ is calculated self--consistently from
a non--linear functional in the autocorrelator $\Phi(t)$ 
of the oscillator variable:
\begin{equation}
m(t)=v_{1}\Phi(t)+v_{2}\Phi(t)^{2}.
\label{e2}
\end{equation}
The coupling vertices $v_i$ in microscopic MCT equations are known regular
functions of 
the  thermodynamic state variables \cite{fuenf}.
In the schematic model the vertices are  considered as coupling constants which
vary smoothly with control variables like  $T$.
Different experimental techniques probe glassy dynamics in a different
manner. Here it is assumed  that a second
correlator 
$\Phi_s(t)$, or  its susceptibility $\chi''_s(\omega)$, describes
the variable probed by  the
depolarized light  scattering spectra as function of frequency $\omega$.
 The correlator $\Phi_s(t)$ 
obeys an equation analogous to eq. \gl{e1} and  couples to the structural
relaxation  
in its friction function via
\begin{equation}
m_{s}(t)=v_{s}\Phi(t)\Phi_{s}(t).
\label{e3}
\end{equation}
Thus, the model is specified by 4 
oscillator parameters ($\Omega$, $\nu$, $\Omega_s$, $\nu_s$)
quantifying the  microscopic excitations in the THz regime which will be taken
as $T$--independent, 
and by 3 equilibrium state control parameters ($v_1$, $v_2$,
$v_s$) whose regular $T$--dependence drives the  spectacular changes of
the  spectra. 


\section{Results}
\noindent
Figure \ref{fig1} reproduces the susceptibility spectra measured by 
depolarized light scattering in OTP \cite{eins} and the fits using 
$\chi''_s(\omega)$ from the schematic model.
The trivial temperature variations of the data due to the Bose factor had been
eliminated already in ref. \cite{eins}.
The spectra and corresponding fits cover  frequency  windows
larger than 3 decades and temperatures from above the melting temperature 
 $T_m=$ 329K
down to below the glass transition temperature $T_g=$ 244K.
 The $\alpha$--relaxation
peak, observed at high temperatures, moves out of the experimental window below
$T=310$K. The  excitations around 1 THz show temperature dependence.
 Nevertheless, the oscillator parameters of the fit
have been used as $T$--independent:
$\Omega/2\pi=1.81$THz, $\nu=0$, $\Omega_s/2\pi=0.14$THz and
$\nu_s/2\pi=0.44$THz.
Obviously, the simple model of two microscopic oscillations does not suffice to
fit the high frequency part of the spectra. We have restricted the fit to
frequencies below 400 GHz.

The temperature change of the theoretical spectra is caused
by the smooth drifts of the coupling vertices shown in fig. \ref{fig2}.
A rather simple path with a $T$--independent 
 parameter $v_1=0.87$ was chosen.
Coupling constants $v_2$ and $v_s$, which increase smoothly
with decreasing temperature drive the system
through the MCT bifurcation singularity, which is located at
$T_c\approx 280$K. The predicted universal features of the
MCT spectra close to $T_c$ \cite{fuenf} are determined by the exponent
parameter 
$\lambda$, which equals $\lambda=0.735$ for this path.

\section{Discussion}
\noindent
 Figure \ref{fig1} verifies that the model captures
the crossover from oscillatory microscopic dynamics near and above  400 GHz to
relaxational processes for lower frequencies. If a Markovian approximation
to the spectra held below the microscopic excitations, a white noise 
susceptibility, $\chi''_{\rm white}\propto \omega$, should be observed.
Such a white noise spectrum is shown in fig. 1 with slope unity.
Both steeper slopes, indicating
oscillatory dynamics at high frequencies, and shallower slopes, indicating 
non--exponential low--frequency relaxation,
 appear in the spectra and in the fits.
The possibility of choosing temperature independent oscillator
parameters, even though $T$--dependence above 400 GHz is apparent in the data,
stresses  that the mode--coupling effects can cause
noticeable temperature variations of the low--frequency wing of the microscopic
excitations. An anomalous increase of the structural relaxation spectra above
the 
white noise approximation is predicted in asymptotic MCT solutions by the
critical power 
law $\chi''(\omega)\propto\omega^a$. It is exhibited by the theoretical curve
labeled $c$ in fig. \ref{fig1}
and its observation is expected to mark the temperature closest to $T_c$. 
The critical exponent $a$ follows from $\lambda$ and equals $a=0.312$ for OTP.

The  $\alpha$--relaxation peak, which lies in the experimental frequency window
for temperatures around the melting temperature  $T_m$, can be characterized
by a $T$--independent Kohlrausch stretching parameter $\beta=0.79$ \cite{eins}.
The $\alpha$--peak
 is fitted  by the schematic model solution which
asymptotically exhibits a von Schweidler power law wing,
$\chi''(\omega)\propto\omega^{-b}$. The exponent $b=0.582$ also follows
from $\lambda$. 
From the detailed analysis of ref. \cite{eins} one can also conclude, as is
corroborated by our fits, that the enhancement of the spectrum above a white
noise background 
in the minimum region of the spectra cannot be explained by a simple
interpolation between a white noise wing of the 
microscopic spectrum and a stretched $\alpha$--process.

In the solutions of the schematic model shown in fig. \ref{fig1}, the
asymptotic idealized MCT predictions carry corrections due to the microscopic
processes 
and due to finite distances of the control parameters from the bifurcation.
In atomic glass formers, additional thermally
activated processes are present and lead to  corrections to the idealized
MCT picture. In the frequency range of the minimum in
$\chi''(\omega)$, these corrections are understood in
asymptotic expansions \cite{sieben}. The activated processes 
included in the extended MCT explain the existence 
of the $\alpha$--process below $T_c$ and the systematic increase of the spectra
at low frequencies and temperatures above the
 fits with the schematic model.
The analysis of the low temperature spectra with the asymptotic predictions of
the extended MCT 
from ref. \cite{eins} is redone in fig. \ref{fig1} using our slightly different
value of 
$\lambda$ and thus somewhat different control parameters. In this analysis one
further $T$--dependent parameter $\delta$ enters characterizing the
thermally activated 
processes \cite{fuenf,sieben}; it is also shown in fig. \ref{fig2}.
Close to the critical temperature, at $T=280$K,
the schematic model solution and the extended MCT asymptotic fit both exhibit
the critical power law and
overlap for about half a decade in frequency. For lower temperatures,
the corrections included in the schematic model, which mask the critical
power law towards higher frequencies, and the corrections to $\chi''\propto 
\omega^a$ due to the thermally activated processes, which come in from lower
frequencies, limit the window
for the asymptotic results of the idealized MCT.
 
\section{Conclusions}
\noindent
The universal asymptotic results of the MCT predict a sensitive dependence of
the structural dynamics on the experimental control parameters like
temperature. 
If corrections to the asymptotic results are large it is difficult to identify
the predicted scaling laws. In this case 
it would be best to use microscopically formulated MCT equations to
quantitatively calculate non-- asymptotic corrections
\cite{acht}.  As a compromise,  schematic MCT models
may be used to test the applicability of the MCT to more complicated systems,
where microscopic MCT equations are yet unknown. The present study of GHz
spectra of OTP demonstrates this approach and
indicates that all observed features of the structural relaxation of OTP
can be described  by smooth temperature dependence of two model parameters.
Including thermally activated processes below $T_c$ a 
complementary fit using the extended MCT covers the spectra at lower
temperatures  and 
matches to the schematic model fit.
The values of the MCT exponents and of the critical temperature which we find
are 
in reasonable agreement with previous neutron \cite{vier} and light scattering
studies \cite{eins}.

This work was supported by 
NATO Collaborative Research Grant No. CRG-930730 and by
Verbundprojekt BMBF 03-G04TUM.

\begin{figure}[ht]
\caption[Figur eins]{\label{fig1} 
Measured susceptibility spectra for OTP from ref.  \cite{eins} (open circles). 
The thin straight line
indicates a white noise spectrum, $\chi''_{white}(\omega)\propto\omega$. The
full lines are MCT solutions for the schematic model eqs. \gl{e1} to \gl{e3}
 with model parameters described in the text and in fig. \ref{fig2}.
The dashed lines show fits for intermediate frequencies
 with the asymptotic scaling law of the extended MCT which are 
discussed in section 4 and in ref. \cite{eins}.}
\end{figure}

\begin{figure}[ht]
\caption[Figur zwei]{\label{fig2} 
Coupling constants  for the schematic--model fits of fig. \ref{fig1}.
The liquid--glass transitions for the model occur either at the 
dashed 
line joining the $(v_1,v_2)$ points (1,0) and (1,1) or at the full parabola
line joining (1,1) and (0,4).
The diamonds with dots mark the $v_{1}-v_{2}$- pairs, chosen for the
solutions of fig. \ref{fig1}, and the star at 280K
indicates the transition point for
the chosen parameter path. 
The inset shows the fit parameters
$v_{2}$ and $v_{s}$ as functions of temperature, connected by lines as guides
to 
the eye. 
The values of the hopping parameter $\delta$ (using
$t_0=1/\Omega$ \cite{sieben}) obtained from
the extended MCT fits shown in fig. \ref{fig1} are also included in
the inset and are close to the values reported in ref. \cite{eins}.
The critical temperature $T_c=$ 280K is marked by a vertical line.}
\end{figure}

\clearpage
\centerline{
\epsfxsize=15cm
{\epsffile{fig1.ps.bb}}}
\bigskip
\centerline{Singh et al., 
Structural Relaxations... \citefigure{fig1}}
\clearpage
\centerline{
\epsfxsize=15cm
{\epsffile{fig2.ps.bb}}}
\bigskip
\centerline{Singh et al., 
Structural Relaxations... \citefigure{fig2}}
\clearpage
\end{document}